\documentclass[preprint,endfloats,eqsecnum,draft,intlimits,amsmath,amssymb]{revtex4}
\usepackage{bm}
\usepackage[final]{graphicx}
\usepackage{longtable}
\newcommand{\be}{\begin{equation}}
\newcommand{\ee}{\end{equation}}
\newcommand{\bea}{\begin{eqnarray}}
\newcommand{\eea}{\end{eqnarray}}

\newcommand{\eqan}[1]{\begin{eqnarray*}#1\end{eqnarray}}

\begin{document}

\title{Adsorption of polymer chains on structured surface: field theoretical approach}

\author{Z.Usatenko}
 \affiliation{Institute for Condensed Matter
Physics, National Academy of Sciences of Ukraine, 79011 Lviv,
Ukraine}

\vspace{0.1cm}
\date{\today}

\begin{abstract}

Taking into account the well known correspondence between the field
theoretical $O(n)$-vector model in the limit $n\to 0$ and the
behavior of long-flexible polymer chains in a good solvent the
investigation of ideal polymer chains adsorption onto structured
surface like as a chemical step (where one part of a surface is
repulsive for polymers and other part is at the adsorption
threshold) was performed. The two-point correlation function of
ideal polymer chain in the half - space bounded by structured
surface with different adsorption energies $c_{1}$ and $c_{2}$ (with
$c_{1}\neq c_{2}$) and the "closest form" for the free propagator of
the model were obtained in analytical form. Besides, the force which
ideal polymer chain with free end exerts on the structured surface,
when the other end is fixed at the surface, was calculated. The
obtained results indicate that the process of homopolymer adsorption
onto structured surfaces should be described by different scaling
laws than universal scaling laws predicted in the literature for
homopolymer adsorption on homogeneous surfaces.

\end{abstract}
\vspace{0.2cm} 

\maketitle

\renewcommand{\theequation}{\arabic{section}.\arabic{equation}}
\section{Introduction}\setcounter{equation}{0}

Adsorption of polymers on the surface has been studied intensively
during last decades because of its rich physics and wide practical
applications such as adhesion, lithography, chromatography, etc. The
early studies mostly were focused on the investigation of the
polymer adsorption onto physically and chemically homogeneous
surfaces \cite{Rubin65,deGennes69,deGennes76,EKB82,E93,U06}. From
another side, during years intensive investigations were devoted to
the problem of adsorption on heterogeneous surfaces both for
homopolymers and heteropolymers such as block copolymers and
periodic or random copolymers. The problem of adsorption from the
solution of polymers onto chemically heterogeneous surfaces using an
analytical self-consistent field theory was investigated by Odijik
\cite{Odijik90} and by Andelman and Joanny in both cases where the
heterogeneity was quenched or annealed
\cite{Andelman-Joanny91,Andelman-Joanny93}. They found that the
heterogeneity of the adsorbing surface enhanced adsorption.
Sebastian and Sumithra \cite{Sumithra1,Sumithra2} analyzed the
influence of surface randomness on the conformation properties of
the adsorbed Gaussian polymer chains using a generalization of de
Gennes's approach \cite{deGennes69} to the case of polymer
adsorption on random surface with taking into account replica trick
and path integral method. It should be mentioned, that one of our
previous works \cite{US07} was connected with description in the
framework of the massive field theory approach of the influence of
the different kinds of surface and near the surface disorder on the
process of homopolymer adsorption on the surface.

 The series of papers were devoted to investigation of
the heteropolymers adsorption on heterogeneous surfaces
\cite{Chakraborty96,Chakraborty98,Chakraborty01, Srebnik00}. Using
different analytical approaches and Monte Carlo simulations they
found that, upon increasing the strength of the interactions, the
heteropolymers adsorption on heterogeneous surfaces is followed by a
second sharp transition, where the polymers freeze into
conformations in such a way that they match the surface pattern.
Such two-step adsorption process describes the physics of protein
(DNA) recognition, where the protein slides on the DNA before
finding its specific docking site.

 During last years a topic of great interest was investigation of copolymers
 adsorption on patterned surfaces in accordance with their importance for
applications in nanotechnology and for the design of novel blood
contacting materials for medical implants and bioaffinity sensors
\cite{Kenausis00,Huang01}. In the framework of the three dimensional
self-consistent field theory was found that the copolymers not only
recognize the patterns, but also propagate the pattern from the
surface to the bulk \cite{Genzer01,Genzer04,Jayaraman05}. Besides,
the phenomena of polymer recognition by multifunctional surfaces
have been studied in a series of works by experiments
\cite{Peppas02,Haupt02}. In work of Tsori and Andelman
\cite{Tsori00} a Ginzburg-Landau free energy was used in order to
investigate the morphology of diblock copolymers in the vicinity of
flat, chemically patterned surfaces. An investigation of the
recognition of patterned substrates by heteropolymer chains was
carried out by Kriksin et al. \cite{Kriksin05,Kriksin06} in the
framework of a "minimal" statistical mechanical model. The density
profiles of diblock copolymers near patterned surfaces covered with
parallel chemically heterogeneous stripes were investigated by
Petera and Muthukumar \cite{Petera98} and by Balazs et
al.\cite{Balazs98}. The adsorption of random heteropolymers in a bad
solvent on patterned surfaces was discussed by Lee and Vilgis in the
framework of a variational approach \cite{Vilgis03}. Most recently
the Monte Carlo simulations performed by Sumithra and Straube
\cite{Sumithra06,Sumithra07} on the adsorption of diblock copolymers
on stripe-patterned surfaces confirmed the previous predictions
obtained by Kriksin et al.\cite{Kriksin05,Kriksin06} that the
process of polymer recognition occurs in two steps characterized by
two transitions: adsorption and freezing transitions, respectively.
However, the detailed analysis showed that the scaling exponents for
the parallel and perpendicular components of the radius of gyration
for diblock copolymer adsorbed on a patterned surface are different
from the scaling exponents describing adsorption of homopolymer on a
homogeneous surface.

 In
\cite{Nath99} and \cite{Sokolowski06} a density functional theory
(DFT) was used for study the ordering of block copolymers near
patterned surfaces and for description of adsorption in systems in
which selected segments of chain molecules can be bound with
functional groups attached to the surface, respectively. In
\cite{Chen06} DFT was applied for the recognition of homopolymer at
nanopatterned surface and was assumed that the segment of the
polymer can recognize one type of stripe and has no interaction with
the other type of stripe.

 As it is known \cite{DD81,D86}, boundaries, which became important at
investigation of the confined systems, induce deviations from the
bulk behavior. The boundary conditions applied to the system
determine the surface universality class, to which the system under
investigation belongs. As it was shown in \cite{D86}, from the point
of view of renormalization group theory it is sufficient to describe
the presence of the substrate by a surface field $h_{1}$ and the
so-called surface enhancement $c$, with $c\sim c_{0}-c_{sp}$ where
$c_{0}$ is the fixed point corresponding to the special transition
located at $c_{0}=c_{sp}$. Thus, the special transition occurs at
$c=0$ in the absence of external fields. The set of papers
\cite{Tsori00,Sprenger-Dietrich05,Gambassi-Dietrich11} were focused
on the crossover from ordinary ($|h_{1}|=0, c>0$) to so-called
normal ($|h_{1}|>0,c=0$) and extraordinary surface universality
classes ($|h_{1}|=0,c<0$), respectively. Less attention was paid to
investigation of adsorption - desorption process of homopolymers at
structured substrates where one type of stripes is repulsive for
polymers ({\it i.e.} is at ordinary surface universality class,
$h_{1}=0, c>0$) and the other type of stripes is inert ({\it i.e.}
is at special surface universality class $h_{1}=0,c=0$ or shortly
speaking is at the adsorption threshold). In accordance with it the
present paper tries to fulfill the gap in the field and is devoted
to investigation of homopolymer adsorption onto {\it flat structured
surface} like as a chemical step, where one part of a surface is
repulsive for polymers and other part is at the adsorption
threshold. Investigation of this type can be generalized in future
for more complicated cases of structured surfaces such as chemical
stripe or periodically structured surface, etc.

The main goal of the present paper is to obtain in an analytical
form the free propagator for such systems with structured surface
and to calculate the force which
 ideal polymer chain with free end exerts on the structured surface, when the other
end is fixed on the surface. The knowledge of the free propagator
for such class of systems is important because it is the
zeroth-order approximation in a systematic Feynman graph expansion
on which the $\epsilon=4-d$ - expansion and massive field theory
approach at fixed space dimensions $d<4$ are based. The higher
orders in the Feynman graph expansion require taking into account
the contribution from the excluded volume interaction.
\section{The model}
In our investigations we consider a dilute polymer solution, where
different polymer chains do not overlap and the behavior of such
polymer solution can be described by a single polymer chain. As it
is known, the single polymer chain can be modeled by the model of
random walk and this describes the ideal polymer chain in
$\theta$-solvent or self-avoiding walk for real polymer chain with
excluded volume interactions (EVI) for temperatures above the
$\theta$-point. Taking into account the polymer-magnet analogy
developed by \cite{deGennes79}, their scaling properties in the
limit of an infinite number of steps $N$ may be derived by a formal
$n \to 0$ limit of the field theoretical $\phi^4$ $O(n)$- vector
model at its critical point, where $1/N$ plays the role of a
critical parameter  analogous to the reduced  critical temperature
in magnetic systems. Besides, as it was noted by de Gennes
\cite{deGennes76} and by Barber et al. \cite{Barber}, there is a
formal analogy between the polymer adsorption problem and the
equivalent problem of critical phenomena in the semi-infinite
$|\phi|^4$ $n$-vector model of a magnet with a free surface
\cite{DD81,D86}. The deviation from the adsorption
 threshold $(c\propto(T-T_a)/T_a)$ (where $T_{a}$ is adsorption temperature)
 changes sign at the transition between the adsorbed ($c<0$) and the
nonadsorbed state ($c>0$) and it plays the role of a second critical
parameter. The value $c$ corresponds to the adsorption energy
 divided by $k_{B}T$ (or the surface enhancement in field theoretical
 treatment). The adsorption threshold for long-flexible
 infinite polymer chains, where $1/N\to 0$ and $c\to 0$ is a multicritical phenomenon.

The effective Ginzburg-Landau Hamiltonian describing the system of
dilute polymer solution is:
 \be {\cal H}[{\vec \phi}] =\int
d^{d}x\bigg\lbrace \frac{1}{2} \left( \nabla{\vec{\phi}} \right)^{2}
+\frac{{\mu_{0}}^{2}}{2} {\vec{\phi}}^{2} \bigg\rbrace,
\label{hamiltonian}\ee

where ${\vec \phi}({\bf x})$ is an $n$-vector field with the
components $\phi_i(x)$, $i=1,...,n$ and  ${\bf{x}}=({\bf r},z)$ is
$d$ dimensional vector, $\mu_0$ is the "bare mass". The present
study is devoted to the investigation of the ideal polymer chain
adsorption onto the structured surface. In the case of dilute
polymer solution in semi-infinite space bounded by structured
surface like as a chemical step (see Figure 1, where for simplicity
of presentation we restrict our attention to the three dimensional
space $d=3$), when from $-\infty$ to $0$ in the $x$ direction the
monomer-surface interaction is described by $c_{1_{0}}$ and from $0$
to $\infty$ the monomer-surface interaction is described by
$c_{2_{0}}$ and $c_{1_{0}}\neq c_{2_{0}}$ (as was mentioned before,
$c_{i_{0}}$ with $i=1,2$ are the corresponding adsorption energies
divided by $k_{B}T$ or the bare surface enhancements in field
theoretical treatment), we should take into account in the
Hamiltonian two additional surface terms:
 \be
\frac{c_{1_0}}{2} \int_{-\infty}^{0}dx\int d^{d-2}{\tilde{\bf{r}}}
{\vec{\phi}}^{2}_{c_{1}}({\tilde{\bf{r}}},x,z=0)
+\frac{c_{2_0}}{2}\int_{0}^{+\infty}dx\int d^{d-2}{\tilde{\bf{r}}}
{\vec{\phi}}^2_{c_{2}}({\tilde{\bf{}r}},x,z=0)\label{struct},\ee
where ${\tilde{\bf{r}}}$ is $d-2$ dimensional vector. The presence
of quadratic surface terms assumes that in $z$ direction the
symmetry of $O(n)$- vector model is broken.
\begin{figure}[ht!]
\begin{center}
\includegraphics[width=7.0cm]{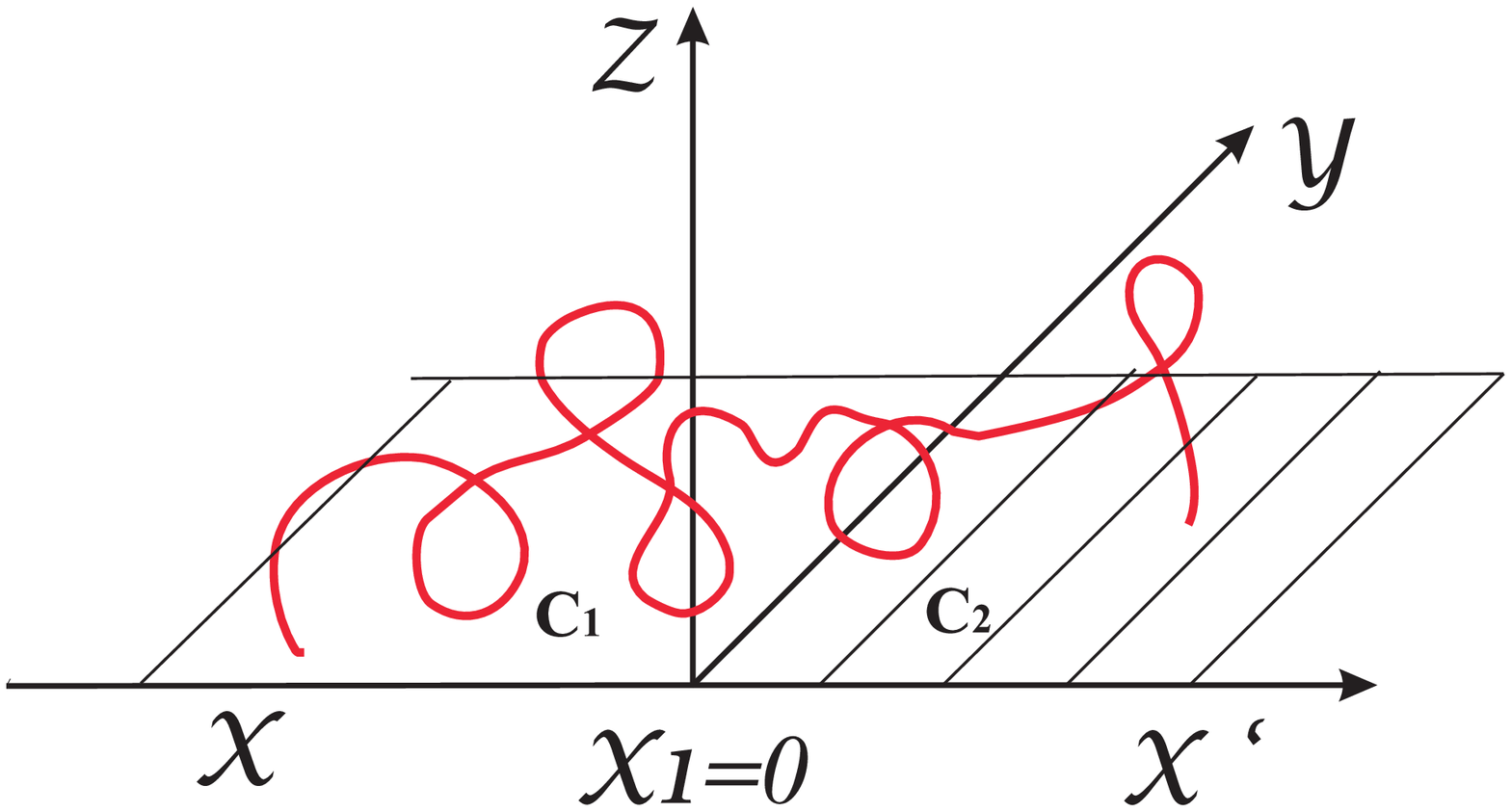} \caption{Adsorption of ideal polymer chain onto the structured surface
with $c_{1_{0}}\neq c_{2_{0}}$.} \label{Fig:1}
\end{center}
\end{figure}
 The interaction between the polymer chain
and the wall is implemented by the different boundary conditions. In
the present paper we are interested in investigation of the
situation when from $-\infty$  to $0$ in the $x$ direction wall is
repulsive ({\it i.e.} where the segment partition function and thus
the partition function for the whole polymer chain $Z({\bf x},{\bf
x}';N)$ tends to 0 as any segment approaches the surface of the
wall) and from $0$ to $\infty$ wall is inert (or at the adsorption
threshold).
 In this case the Dirichlet and the Neumann boundary conditions takes place on
each piece of wall, respectively (see Fig.1): \be c_{1}\to
+{\infty},\quad c_{2}=0\quad or\quad {\vec \phi}_{c_{1}}({\bf
{r}},0)=0, \quad\frac{\partial{{\vec \phi}_{c_{2}}({\bf
{r}},z)}}{\partial z}|_{z=0}=0\label{DN},\ee where $c_{1}$ and
$c_{2}$ are renormalized surface enhancements. Besides, for
completeness of description we also discuss the reverse case with
Neumann and Dirichlet boundary conditions, respectively. In the last
mentioned case from $-\infty$  to $0$ in the $x$ direction wall is
inert and from $0$ to $\infty$ wall is repulsive.

Taking into account the polymer-magnet analogy \cite{deGennes79},
the partition function $Z_{c_{1_{0}},c_{2_{0}}}({\bf{x}},{\bf{x'}})$
of a single polymer chain with two ends fixed at ${\bf{x}}$ and
${\bf{x'}}$ is connected with the two-point correlation function
$G_{c_{1_{0}},c_{2_{0}}}({\bf{x}},{\bf{x'}})=<{\vec{\phi}({\bf
x})}{\vec{\phi}}({\bf x}')>^{c_{1_{0}},c_{2_{0}}}$  in a
Ginzburg-Landau model via the inverse Laplace transform
$\mu_{0}^{2}\to L_{0}$ \be Z_{c_{1_{0}},c_{2_{0}}}({\bf x},{\bf
x}';N)={\cal IL}_{\mu_{0}^2\to L_{0}}(<{\vec \phi}({\bf
x}){\vec\phi}({\bf x'})>^{c_{1_{0}},c_{2_{0}}}|_{n\to
0})\label{critpoly} \ee in the limit, where the number of components
$n$ tends to zero. The conjugate Laplace variable $L_{0}$ has the
dimension of length squared and is proportional to the total number
of monomers $N$ of the polymer chain: $L_{0}=N{\tilde{l}}^{2}$,
where $\tilde{l}$ is the microscopic length of monomer size. For
ideal polymer chains $N$ equals $<{\bf{R}^2}>$. It should be
mentioned, that the most common parameter in polymer physics to
denote the size of polymer chains which are observable in
experiments is radius of gyration $R_{g}$ \cite{CJ90,E93,Sh98} which
is $R_{g}^{2}=R_{x}^2/2$ for $d=3$ case, where $R_{x}$ is the
projection of the end to end distance ${\bf R}$ onto the direction
of $x$ axis: $R_{x}^{2}=<\frac{\bf{R}^{2}}{d}>$.

\section{Correlation function of ideal polymer chain in the half-space bounded by structured surface with $c_{1}\neq c_{2}$.}

In general, in order to remove UV singularities of the correlation
function $G_{c_{1_{0}},c_{2_{0}}}({\bf{x}},{\bf{x'}})$ located in
bulk or on the surface a mass shift $m_{0}^{2}=m^{2}+\delta m$ and a
surface-enhancement shift is required $ c_{i_{0}}=c_{i}+\delta
c_{i}$ (with $i=1,2$), respectively (see \cite{DSh98}). In the
present order of approximation scheme $\delta c_{i}=0$ and in
accordance with it for ideal polymer chain we can replace
$c_{1_{0}}\to c_{1}$ and $c_{2_{0}}\to c_{2}$. In general case we
can rewrite (\ref{struct}) in the form: \bea \frac{c_{1}}{2}
\int_{-\infty}^{+\infty}dx\int d^{d-2}{\tilde{\bf{r}}}
{\vec{\phi}}^{2}_{c_{1}}({\tilde{\bf{r}}},x,z=0)
+\frac{c_{2}}{2}\int_{0}^{+\infty}dx\int d^{d-2}{\tilde{\bf{}r}}
{\vec{\phi}}^2_{c_{2}}({\tilde{\bf{r}}},x,z=0)\nonumber\\
-\frac{c_{1}}{2}\int_{0}^{+\infty}dx\int d^{d-2}{\tilde{\bf{}r}}
{\vec{\phi}}^2_{c_{1}}({\tilde{\bf{r}}},x,z=0).\label{perep-struct}
\eea If we assume, that
$\frac{1}{2}{\vec{\phi}}^{2}_{c_{1}}=\frac{1}{2}{\vec{\phi}}^{2}_{c_{2}}$
and take into account that in the framework of the present
approximation scheme the deviations $\Delta c= c_{2}-c_{1}$ are
small enough  the correspondent two-point correlation function
$G_{c_{1},c_{2}}$ can be written in the form \bea
G_{c_{1},c_{2}}(x,z;x'z';{\bf{\tilde{p}}})&=&G_{c_{1}}(x,z;x'z';{\bf{\tilde{p}}})\nonumber\\-(c_{2}-c_{1})\int_{0}^{\infty}
G_{c_{1}}(x,z;x_{1},z_{1}=0;{\bf{\tilde{p}}})&&G_{c_{1}}(x_{1},z_{1}=0;x',z';{\bf{\tilde{p}}})+...,\eea
which allows to describe the crossover from special ($c=0$) to
ordinary ($c\to\infty$) surface transition and includes arbitrary
number of surface operator $\frac{1}{2}{\vec{\phi}}^{2}_{c_{i}}$
insertions. In  \cite{DSh98} was assumed, that for small deviation
$\Delta c= c_{2}-c_{1}$ the system is still translationally
invariant in direction of changing from $c_{1}$ to $c_{2}$, which
corresponds to "x" direction in our case. Actually in the present
paper the similar assumption is applied. It should be mentioned,
that such assumption is realistic, because we do not introduce any
surface field  which can destroy translational invariance in the
remaining $d-1$ directions as it was discussed in the previous
papers (see \cite{Tsori00,Petera98} etc.), but take into account
change of the value of adsorption energy from $c_{1}$ to $c_{2}$.
Unfortunately, the approach proposed in \cite {DSh98} does not give
us possibility to distinguish the contribution from different
regions of structured surface. In accordance with it in order to
distinguish the contributions from $c_{1}$ and $c_{2}$ regions in
the present paper is assumed, that surface operator
$\frac{1}{2}{\vec{\phi}}^{2}({\bf{\tilde{r}}_{1}},x_{1},z_{1})$
insertions contain different components for $c_{1}$ and $c_{2}$
regions: $\frac{1}{2}{\vec{\phi}}^{2}_{c_{1}}\neq
\frac{1}{2}{\vec{\phi}}^{2}_{c_{2}}$.  Thus, if we take into account
that one end $(\tilde{{\bf{r}}},x,z)$ of polymer chain can be in
$c_{1}$ region and second end $(\tilde{{\bf{r}}}',x',z')$ in $c_{2}$
region, or two ends can be in $c_{1}$ or $c_{2}$ regions,
respectively, the correspondent correlation function for ideal
polymer chain in the half space restricted by structured surface
with $c_{1}\neq c_{2}$ in the mixed momentum - space representation
is:
\bea G_{c_{1},c_{2}}(x,z;x',z';{\tilde{\bf{p}}})=G_{c_{1}}(x,z;x',z';{\tilde{\bf{p}}})-\nonumber\\
(c_{2}-c_{1})\int_{0}^{\infty} dx_{1} G_{c_{1}}(x,z;
x_{1},z_{1}=0;{\tilde{\bf{p}}})G_{c_{2}}(x_{1},z_{1}=0;x',z';{\tilde{\bf{p}}})+...
,\label{init-corf} \eea where ${\tilde{\bf{p}}}$ is $d-2$
dimensional vector of momentum. As it is easy to see from
Eq.(\ref{init-corf}), the insertion of surface operators
$\frac{1}{2}{\vec{\phi}}_{c_{i}}^{2}(\tilde{\bf{r}},x_{1},z_{1}=0)$
with $i=1,2$ assume that the difference in the monomer-surface
interactions on the both pieces of the wall ($c_{1}\neq c_{2}$) is
restricted to the surface (see Fig.1) and in accordance with it we
have $z_{1}=0$ everywhere.

  The proposed in the present paper approach uses some ideas proposed
by Symanzik \cite{Symanzik81} and, as it was mentioned before, is in
some way similar to the investigation of crossover region between
special and ordinary transition, proposed in \cite{DD81,D86,DSh98},
but allows to distinguish the different values of adsorption
energies $c_{1}$ and $c_{2}$ on flat structured surface with taking
into account that deviations $c_{2}$ from $c_{1}$ should be small
enough.
 The functions $G_{c_{1}}(x,z;
x',z';{\tilde{\bf{p}}})$ and $G_{c_{2}}(x,z;
x',z';{\tilde{\bf{p}}})$ are free propagators in the mixed momentum
- space representation for the system with one surface enhancement
$c_{1}$ or $c_{2}$, respectively. In general we can write: \be
G_{c_{i}}(x,z;x',z';{\tilde{\bf{p}}})=\int_{-\infty}^{+\infty}\frac{d
p_{x}}{2\pi}e^{-ip_{x}(x-x')}G_{c_{i}}(z,z';{\bf{p}}),\ee where
$G_{c_{i}}(z,z';{\bf{p}})$ is the usual free propagator in the mixed
 ${\bf{p}}z$ representation (where ${\bf{p}}$ is $d-1$ dimensional
vector of momentum) for the semi-infinite geometry \cite{DSh98}: \be
G_{c_{i}}(z,z';{\bf{p}})=\frac{1}{2\kappa_{0}}[e^{-\kappa_{0}|z-z'|}-
\frac{(c_{i}-\kappa_{0})}{(c_{i}+\kappa_{0})}e^{-\kappa_{0}(z+z')}],\label{free-usual}
\ee for $i=1,2$ with $\kappa_{0}=\sqrt{p^{2}+\mu^{2}_{0}}$. As it
was mentioned before (see Eq.(\ref{init-corf})) in accordance with
that the difference in the monomer-surface interactions on the both
pieces of the wall ($c_{1}\neq c_{2}$) is restricted to the surface,
i.e. $z_{1}=0$ everywhere, we have that $z\geq z_{1}$ and $z'\geq
z_{1}$, because our polymer ends ${\bf{x}}=(\tilde{{\bf{r}}},x,z)$
and ${\bf{x}}'=(\tilde{{\bf{r}}}',x',z')$ can be everywhere in the
half space restricted by structured surface. In accordance with that
in the case of present model $z\geq z_{1}$,  we obtain the following
relation: \be \frac{\partial}{\partial
z_{1}}[G_{c_{1}}(x,z;x_{1},z_{1};{\tilde{\bf{p}}})]_{|_{z_{1}=0}}=c_{1}G_{c_{1}}(x,z;x_{1},z_{1};{\tilde{\bf{p}}})_{|_{z_{1}=0}}.\label{condition1}\ee
 The present model also assumes that $z'\geq z_{1}$ and in
accordance with it the following relation takes place: \be
\frac{\partial}{\partial
z_{1}}[G_{c_{2}}(x_{1},z_{1};x',z';{\tilde{\bf{p}}})]_{|_{z_{1}=0}}=c_{2}G_{c_{2}}(x_{1},z_{1};x',z';{\tilde{\bf{p}}})_{|_{z_{1}=0}}.\label{condition2}\ee
 Taking into account Eqs.(\ref{condition1})-(\ref{condition2}), our two-point correlation
function in Eq.(\ref{init-corf}) can be written in the form: \bea
G_{c_{1},c_{2}}(x,z;x',z';{\tilde{\bf{p}}})=G_{c_{1}}(x,z;x',z';{\tilde{\bf{p}}})\nonumber\\
+\int_{0}^{\infty}dx_{1}\frac{\partial}{\partial
z_{1}}[G_{c_{1}}(x,z;x_{1},z_{1};{\tilde{\bf{p}}})]_{|_{z_{1}=0}}G_{c_{2}}(x_{1},z_{1}=0;x',z';{\tilde{\bf{p}}})\nonumber\\
-\int_{0}^{\infty}dx_{1}G_{c_{1}}(x,z;x_{1},z_{1}=0;{\tilde{\bf{p}}})\frac{\partial}{\partial
z_{1}}[G_{c_{2}}(x_{1},z_{1};x',z';{\tilde{\bf{p}}})]_{|_{z_{1}=0}}
+...\label{final-corf}
 \eea

 If we assume, that for small deviation $c_{2}$ from $c_{1}$ the system is still
translationally invariant in direction of "x", because we do not
introduce any surface field which can destroy translational
invariance of the system, then in accordance with
Eq.(\ref{final-corf}) the corresponding two-point correlation
function of ideal polymer chain with one end fixed at
${\bf{x}}=({\bf{r}},z)$ and the other end in layer $z'$ in the mixed
${\bf{p}}z$ representation in the half - space bounded by structured
surface with two different adsorption energies $c_{1}$ and $c_{2}$
in general case will be:
 \be
G_{c_{1},c_{2}}(z,z';{\bf{p}})=G_{c_{1}}(z,z';{\bf{p}})
+e^{-\kappa(z+z')}\frac{(c_{1}-c_{2})}{(c_{1}+\kappa)(c_{2}+\kappa)}+...,\label{finlast_corf}
\ee with $\kappa=\sqrt{p^2+\mu^2}$. It should be mentioned that in
the present order of approximation scheme we have
$\mu^{2}_{0}\to\mu^{2}$. In the case, when $c_{1}=c_{2}$ from
Eq.(\ref{finlast_corf}) the usual free propagator
Eq.(\ref{free-usual}) (see \cite{DSh98},\cite{Symanzik81}) can be
obtained.

For convenience of representation we can rewrite
(\ref{finlast_corf}) in the so-named the "closest form" (or the
short form): \be
G_{c_{1},c_{2}}(z,z';{\bf{p}})=\frac{1}{G_{c_{1}^{-1}}(z,z';{\bf{p}})-\Sigma({\bf{p}})},\label{fr-prop}\ee
where the corresponding mass operator is:
\be\Sigma({\bf{p}})=G_{c_{1}}^{-2}(z,z';{\bf{p}})e^{-\kappa(z+z')}
\frac{(c_{1}-c_{2})}{(c_{1}+\kappa)(c_{2}+\kappa)}
\label{mass-operator}. \ee

\section{Partition function}

The knowledge of two-point correlation function (see
Eq.(\ref{finlast_corf})) allows easily to obtain via
Eq.(\ref{critpoly}) the corresponding partition function
$Z_{c_{1},c_{2}}({\bf x},{\bf x}';L_{0})$ of ideal polymer chain
with two ends fixed at ${\bf{x}}=({\bf{r}},z)$ and
${\bf{x}}'=({\bf{r}}',z')$ in the half space restricted by
structured surface like as a chemical step. Taking into account
Eq.(\ref{critpoly}) and Eq.(\ref{finlast_corf}), we can obtain, for
example, the partition function of ideal polymer chain with one end
fixed at ${\bf{x}}=({\bf{r}},z)$  and other end free in
semi-infinite space $z'>0$ in the case when surface is repulsive
from $-\infty$ to 0 and inert  from 0 to $+\infty$: \be
Z_{c_{1},c_{2}}(z;L_{0})=\int_{0}^{\infty} dz'
Z_{c_{1},c_{2}}(z,z';L_{0})\approx 1,\label{partf-inert} \ee where
$Z_{c_{1},c_{2}}(z,z';L_{0})={\cal IL}_{\mu_{0}^2\to
L_{0}}G_{c_{1},c_{2}}(z,z';{\bf{p}})|_{n\to 0}.$ It should be
mentioned that here we assume that fixed end of polymer is in
repulsive region and the analytical continuation to the region
$c_{1}\to \infty$ and $c_{2}\to 0$ was performed.  In the case, when
one end of polymer chain is fixed directly at the surface, {\it
i.e.} $z=0$, and other end is free, the corresponding partition
function $Z_{c_{1},c_{2}}(0;L_{0})$ in the above mentioned case
$c_{1}\to \infty$ and $c_{2}\to 0$ is also equal to 1.

 The partition function of ideal polymer chain with one fixed end at
${\bf{x}}=({\bf{r}},z)$ and other end free in semi-infinite space
$z'>0$ in the case when surface is at the beginning inert from
$-\infty$ to 0 and later is repulsive from 0 to $+\infty$ will be :
\be
Z_{c_{1},c_{2}}(z;L_{0})=Erf(\frac{z}{\sqrt{2}R_{x}}),\label{partf-repuls}
\ee where the analytical continuation to the region $c_{1}\to 0$ and
$c_{2}\to \infty$ was assumed and suggestion that fixed end of
polymer is in inert region was taken into account.

\section{Calculation of the force}

Let's consider the force per unit area in the direction
perpendicular to the surface which ideal polymer chain with one free
end exerts on structured surface, when the other end is fixed on the
surface. In accordance with Eq.(\ref{critpoly}), the corresponding
force is: \be
\frac{f_{c_{1},c_{2}}(z')}{k_{B}T}=\frac{\partial}{\partial z'}ln
{\cal IL}\int d^{d-1}{\bf{r}}' G_{c_{1},c_{2}}({\bf{r}},z=0;
{\bf{r}}',z'),\label{force1}\ee
 where ${\cal IL}$ is an inverse Laplace transform and the expression under $ln$ is the
 partition function of ideal polymer chain
 with one end ${\bf{x}}=({\bf{r}},z=0)$ fixed on the surface and with other end in the layer $z'$.
 The force Eq.(\ref{force1}) depends on the energy of adsorption,
 because $c_{1}$  and $c_{2}$ corresponds to the adsorption energy
 divided by $k_{B}T$.
 This force is analogous to the well known Pincus force
 \cite{deGennes79,BMRV09}, which is necessary to apply in order to detach
 single polymer chain from homogeneous surface. Taking into account
Eq.(\ref{finlast_corf}), the corresponding force in the case when
surface is repulsive from $-\infty$ to 0 and inert from 0 to
$+\infty$ is: \be \frac{f_{c_{1},c_{2}}(z')}{k_{B}T}\approx
\frac{1}{z'}(1-\frac{(z')^{2}}{R_{x}^{2}}).\label{force-repulsive}\ee

 It should be mentioned, that here the analytical
continuation to the region, where $c_{1}\to \infty$ and $c_{2}\to 0$
was performed. The resulting force in this case is repulsive (see
Fig.2).
\begin{figure}[ht!]
\begin{center}
\includegraphics[width=7.0cm]{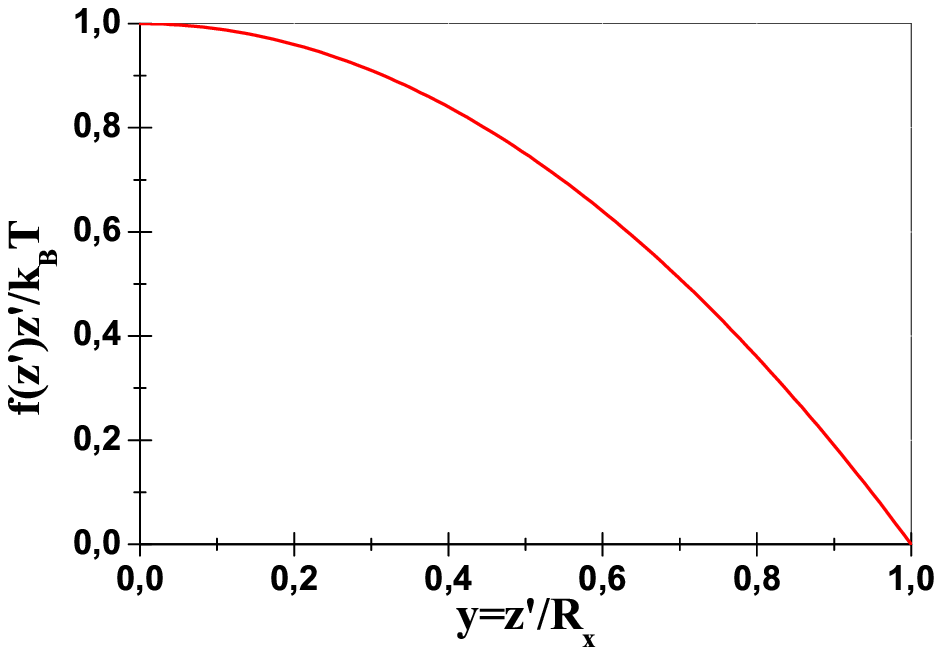}
\caption{The dimensionless value of the force per unit area in the
direction perpendicular to the surface in the form $f(z')z'/k_{B}T$
for ideal polymer chain with one fixed end on the surface and other
end free $z'>0$ near structured surface with $c_{1}\to\infty$,
$c_{2}\to 0$. The force is repulsive.}
\label{fig:struct-force-repulsive}
\end{center}
\end{figure}

The force, which is necessary apply to ideal polymer chain with one
fixed end at the surface and the other end free in the half-space
$z'>0$ in the case when surface is at the beginning inert from
$-\infty$ to 0 and later is repulsive from 0 to $+\infty$, tends to
infinity. It assumes, that it is very difficult to detach adsorbed
polymer from the surface.

Thus, depending on that in which region (repulsive or inert) of
structured substrate of the type Eq.(\ref{struct}) the free end of
ideal polymer chain is localized, the total force per unit area in
the direction perpendicular to the surface which ideal polymer chain
exerts on the surface can be repulsive or attractive, respectively.
The position of free end of ideal polymer chain which is anchored by
other end to the surface has decisive influence on its critical
behavior near structured surfaces as it was also confirmed during
our calculations of the corresponding partition functions.

\section{Concluding remarks and summary}

The investigation of the process of ideal homopolymer adsorption -
desorption onto structured surface like as a chemical step (where
one part of a surface is repulsive for polymers and other part is at
the adsorption threshold) was performed in the framework of the
field theoretical approach. The presented in this paper approach
uses some ideas proposed by Symanzik \cite{Symanzik81} and by Diehl
and Shpot \cite{DSh98} for the investigation of crossover region
between special and ordinary transition, but allows to distinguish
the values of adsorption energies (or surface enhancement in field
theoretical treatment) $c_{1}$ and $c_{2}$ on structured surface
with taking into account that deviations $c_{2}$ from $c_{1}$ are
small enough. The main obtained results of the present paper are the
following.

(1)The two-point correlation function of ideal polymer chain in the
mixed ${\bf{p}}z$ representation in the half - space bounded by
structured surface with two different adsorption energies $c_{1}$
and $c_{2}$ (see Eq.(\ref{finlast_corf})) and the "closest form" for
the free propagator of the model (\ref{hamiltonian}) with
(\ref{struct}) were obtained (see Eq.(\ref{fr-prop})) for the first
time in analytical form. The knowledge of the analytical form for
the free propagator for such class of systems is important because
it is the zeroth-order approximation in a systematic Feynman graph
expansion for real polymer chains with EVI on which the
$\epsilon=4-d$ - expansion and massive field theory approach at
fixed space dimensions $d<4$ are based.

 (2) The force per unit area in the direction
perpendicular to the surface which ideal polymer chain with free end
in semi-infinite space $z'>0$ exerts on structured surface, when the
other end is fixed at the surface ${\bf{x}}=({\bf{r}},z=0)$ in the
case when surface is at the beginning repulsive from $-\infty$ to 0
and later is inert from 0 to $+\infty$ was calculated. The obtained
result indicates that in this case resulting force is repulsive, as
it is possible to see on Fig.2.

(3) Depending on that in which region (repulsive or inert) of
structured substrate of the type Eq.(\ref{struct}) the free end of
ideal polymer chain is localized, the total force per unit area in
the direction perpendicular to the surface which anchored polymer
chain with free end exerts on structured surface can be repulsive or
attractive, respectively.

 (4)The obtained
results indicate that the process of homopolymer adsorption onto
structured surfaces
 should be described by different scaling laws than universal
scaling laws predicted in the literature for homopolymer adsorption
on homogeneous surfaces, because the above mentioned values for the
correlation function, the partition function and for the force
depend not only on one adsorption energy, but from two different
adsorption energies $c_{1}$ and $c_{2}$ (with $c_{1}\neq c_{2}$).
The detailed scaling analysis of homopolymer adsorption onto
structured surfaces are currently in progress.

The present study create basis for further analytical investigations
of critical behavior of real polymer solutions with excluded volume
interactions restricted by structured substrates with more
complicated architecture which will be the subject of our future
investigations. Besides, the proposed in the present paper
investigations are basis for creation in the future of analytical
approach for description of polymer solution in confined geometries
of two nano-structured surfaces and open wide possibilities for
creation of new generation of nano- and micro-mechanical devices
with low static friction, similarly as it was discussed recently for
critical binary fluid mixtures in confined geometries
\cite{DBNat08}.

\section*{Acknowledgments}
Z.U. gratefully acknowledge fruitful discussions with H.W.Diehl and
M.Shpot and thanks for the hospitality at the Duisburg-Essen
University. This work in part was supported by grant from the DAAD
Foundation.

\end{document}